\documentclass[5p,twocolumn,11pt]{elsarticle}

\usepackage{hyperref}
\usepackage{bm}
\usepackage{amssymb} 

\journal{arXiv}

\begin{document}

\begin{frontmatter}

\title{Theoretical insight into the thermoelectric behavior of
	tri-nuclear metal-string complexes laced with gold
	nanoelectrodes: a \textit{first-principles} study}

\author[1,2]{Talem Rebeda Roy}

\author[1,2]{Arijit Sen\corref{cor1}%
}
\ead{arijits@srmist.edu.in}

\cortext[cor1]{Corresponding author}

\address[1]{SRM Research Institute, SRM Institute of Science and Technology, Chennai 603203, India}
\address[2]{Department of Physics \& Nanotechnology, SRMIST, Chennai 603203, India}

\begin{abstract}
Metal-string complexes in the quasi-1D framework may play an  important role in molecular electronics  by serving not only as nanoscale interconnects but also as active functional elements for nanoelectronic devices. However, because of the potential volumetric heat generation across such nanojunctions, the circuit stability  becomes often a major concern, which necessitates to study the heat transport properties at the molecular-scale. Here we report the thermoelectric behavior of various tr-nuclear  metal-string complexes, $[M-M-M](dpa)_4(NCS)_2$ for $M \in \{Cr,Ru\}$, bridging   Au(111) nanowires as nanoelectrodes. Based on our  charge transport analysis from \textit{first-principles}, we find that the dominant transmission peaks tend to move away from the Fermi level upon systematic rutheniation in chromium-based metal-string complexes due mainly to the coupling of $\pi^{*}$  orbitals from Ru and  $\sigma_{nb}$ orbitals from Cr. Such type of a metal-string junction can also exhibit strong Coulomb interaction so that its thermoelectric behavior begins to deviate from the Wiedemann-Franz law. Our results further suggest that metal-string complexes can render better thermoelectric devices especially at the molecular-scale with the thermopower as high as 172 $\mu V/K$ at 300 K. Considering the contributions from both electrons and phonons, even a high \textit{figure of merit} of $ZT \sim 2$ may be attained for Cr-Cr-Cr based metal-string molecular junctions at room temperature. Resonant enhancement in the thermoelectric efficiency appears to occur in such systems through alteration of inter-dot electrostatic interactions, which can be controlled by incorporating Cr and Ru atoms in such tri-nuclear metal-string complexes.

\end{abstract}

\begin{keyword}
\texttt{Metal-string complex, Molecular electronics, Breit–Wigner resonance, NEGF-DFT, Thermopower, ZT} 
\end{keyword}

\end{frontmatter}

\section{Introduction}

Finding a possible way out for accelerating the heat dissipation remains often a major hurdle in the development of nanoscale devices, whose optimal performance gets degraded due especially to overheating. A nanodevice usually operates within the thermal broadening regime as its size gets drastically shrunk beyond the phase coherence  length\cite{Miao2018,Klockner2017n,famili2018,Huang2017}, which is often the case especially with single-molecule devices. The temperature gradient ($\Delta T$) due to phonon excitation gives rise to a open-circuit bias ($\Delta V$) across the junction, as defined by thermopower ($S=\Delta V/\Delta T$) while the efficiency of heat conversion is quantified by a dimensionless quantity, called \textit{figure of merit}\cite{Houten1992,Nolas2001}, as given by $\left|ZT\right|=\sigma S^{2}T/(\kappa_{el}+\kappa_{ph})$, where $\sigma$ ($\kappa$) refers to electronic (thermal) conductance. Study of thermoelectric behavior in metal-molecule-metal junctions deems important since such junctions are often subject to strong molecular vibrations\cite{Mani2011,Guo2013}.

During charge transport, the electrons are likely to interact with the molecules in the tunneling regime leading to phonon emission. In order to enhance the on-chip cooling of molecular devices all these thermal agitations need to be efficiently harvested into workable electrical energy, which can be obtained through favorable combinations of molecules and electrodes constituting  molecular junctions. With the aid of a mechanically controlled break junction (MCBJ) set up, Reddy \textit{et al}\cite{Reddy1568} observed a reduction in the thermopower with the increase in the molecular length. Besides, Huang \textit{et al}\cite{Huang2017} reported  $ZT=0.23\pm0.03$ for a small organo-molecular junction through fine tuning of its conjugated backbone. A theoretical prediction by Mohammed \textit{et al}\cite{Mohammed2017} showed egde-over-edge ZnP dimers with high thermopower and reduced thermal conductance, where the rotation in locked pyridyl ring helped to avert reduction in conductance. Subsequent efforts have been made to develop advanced low-dimensional thermoelectric nanodevices (\textit{e.g}. 1D and 2D nanostructures) which are expected to display efficient thermoelectric properties as compared to their bulk forms. Such quantum-confined nanostructures can serve as better phonon waveguides due to the prevalence of prominent surface phonon scattering\cite{sevinccli2013,Nikolic2012,Dresselhaus1993}, and be potentially utilized to manipulate themoelectric properties for the fabrication of energy conversion devices\cite{Dresselhaus1993}. For example, Hochbaum \textit{et al}\cite{hochbaum2008} and Boukai \textit{et al}\cite{boukai2008} in their respective works reported Si nanowires to give better thermoelectric response than what could be achieved with bulk silicon.

Simplistically, a molecular junction may often be captured by a double quantum dot (DQD) model \cite{Wierzbicki2011, Sen2010},  although such junctions have larger energy level spacing than those present in quantum dots\cite{Mani2011}. While the electronic conductance depends largely on the degree of delocalization along with the electrode-molecule coupling, the thermopower can effectively portray the built-in potential developed across junctions whenever a certain temperature gradient is maintained. In view of this, we investigate here the thermoelectric response of molecular junctions having metal-string complexes $[M-M-M](dpa)_4(NCS)_2$ for $M \equiv \{Cr,Ru\}$, as active elements coupled to Au(111) nanowires as electrodes. All these  metal-strings are capped with $isothiocyanate$ in the form of axial ligands, at both ends which eventually bridged the gold nanoelectrodes through thiol linkers. Since each metal-string under study consists of chromium or ruthenium as metal centers, the high density of electrons around the Fermi level is expected to  enhance the carrier transport\cite{Sen2010,Pontes2008}. Furthermore, the short bond lengths between the adjacent metal centers induces strong interactions in the molecular moieties to help form potentially stable junctions and facilitate charge transport across it.

\section{Computational methods}

In the present study on single-molecule junctions, the central region contained metal-string complexes sandwiched between six layers of Au(111) nanowires. All the device configurations, as shown in Fig. 1, were optimized  by utilizing the density functional theory (DFT) based on the atomic basis set of double-$\zeta$ with polarization (DZP) orbitals in the LCAO (\textit{i.e.} linear combination of atomic orbitals) approach as implemented in the SIESTA package\cite{SIESTA2002}. The exchange and correlation effects were treated considering the Perdew-Burke-Ernzerhof (PBE) functional within the generalized gradient approximation (GGA)\cite{Perdew1996}. Besides, \textit{k}-point sampling of $1\times 1\times 150$ was used with an energy cut-off of 200 Ry. The residual forces were made lower than 0.02 eV/$\AA$ by keeping as many as six electrode layers in the scattering region fixed so as to screen out the perturbation caused by the electrode-molecule interaction, while each electrode possesses an effective potential similar to that of its periodic part. The intra-string bond lengths were found to be in good agreement with the values reported by Huang Hsiao \textit{et al}\cite{Hsiao2008} and Niskanen \textit{et al}\cite{Niskanen2012}. 

Following the structural relaxation of various molecular junctions under study, the charge transport calculations were carried out by utilizing the non-equilibrium Green's function (NEGF) formalism\cite{Taylor2001,Brandbyge2002,Sen2010,asen2013} as implemented in the Atomistix Toolkit (ATK) simulation tool \cite{Atk2015} on top of DFT,  to take care of the open boundary conditions at the junction level in a possible realistic manner. As mentioned before, a two-probe single-molecule junction possesses extended electrodes within the central region acting as a buffer layer. The same forms of exchange-correlation and basis set were utilized as in the case of device relaxation. Similarly, the effect of ion cores on the valence electrons were expounded by making use of norm-conserving Troullier-Martins pseudopotentials\cite{TM}. In each case, self-consistency was achieved by considering 400 \textit{k}-points along the transport direction. 

In the coherent tunneling limit, the energy \textit{E}  transmission function \textit{T(E)} can be obtained from the generalized Landauer theory\cite{Taylor2001,Brandbyge2002,Landauer1957} as

\begin{equation}
T(E)=Tr[\Gamma_{L}(E)G^{r}(E)\Gamma_{R}(E)G^{a}(E)],
\end{equation}  

where the level broadening due to left(right) electrodes is given by $\Gamma_{L}(\Gamma_{R})$ while $G^{r}(G^{a})$ denotes the unperturbed retarded (advanced) Green's functions. The Seebeck coefficient or thermopower can thus be derived by utilizing the Mott's formula as\cite{Jason1980,Sivan1986,Esfarjani2006}
\begin{equation}
S=\frac{L_1(\mu)}{eTL_0(\mu)}
\end{equation}

where,

\begin{equation}
L_m(\mu)=\frac{2}{h}\int_{-\infty}^{\infty}T(E)(E-\mu)^m(-\frac{\partial f(E,\mu)}{\partial E})dE,
\end{equation}
with the integer \textit{m} going from 0 to 2, $h$ being the Planck's constant, \textit{f} the Fermi functions, and $\mu$ the chemical potential of the electrodes. This way, the electronic and thermal conductance can be calculated  as\cite{Brandbyge2002,Landauer1957}

\begin{equation}
\sigma=e^2L_0
\end{equation}

\begin{equation}
\kappa_{el}=\frac{1}{T}(L_2(\mu)-\frac{[L_1(\mu)]^2}{L_0(\mu)})
\end{equation}

The transmission probability of phonons, $T_{ph}(\omega)$, for the junctions were computed from the dynamical matrix ($D_{i\alpha, j\beta}$) based on the \textit{first-principles} approach  as implemented in the Gollum transport   code\cite{Ferrer2014,Sadeghi2015,Hou2019} so that
\begin{equation}
D_{i\alpha, j\beta}=K_{i\alpha, j\beta}/\sqrt{m_im_j}
\end{equation}
where
\begin{equation}
K_{i\alpha, j\beta}=-\frac{F_{j\beta}(Q_i\alpha)-F_{j\beta}(-Q_i\alpha)}{2Q_{i\alpha}}
\end{equation}
Here, $ K$ is the force constant matrix, $F_{j\beta}(Q_i\alpha)$ signify the forces on each of the $j^{th}$ atom along the $\beta$ direction with respect to the displacement of the $i^{th}$ atom along the $\alpha$ direction, and $m_i$($m_j$) denotes the mass of $i^{th}$($j^{th}$) atom. The phononic contributions towards the thermal conductance is thus obtained as

\begin{equation}
\kappa_{ph}=\int_{0}^{\infty}\frac{\hbar\omega}{2\pi}T_{ph}(\omega)\frac{\partial f_{BE}(\omega,T)}{\partial T}d\omega,
\end{equation}

where $f_{BE}$ refers to the Bose-Einstein distribution function, given as $1/[e^{\hbar\omega/K_BT}-1]$. 

\section{Results and discussion}

A schematic of the DQD model is shown in Fig. 1(a) where the two quantum dots with energy difference $\Delta E$ are connected through the tunneling barrier to a cold and hot electron reservoirs. The left(right) electrode serves as a cold(hot) electron reservoir having the electrochemical potential $\mu_{C} (\mu_{H})$ giving rise to a open-circuit bias energy of $eV_{0}$. Here, the energy difference of the quantum dots are assumed to be much larger than the thermal energy \textit{i.e.}, $\Delta E\gg K_BT$, where $K_B$ is the Boltzmann constant and $T$ is the average temperature of the reservoirs. An electron with the resonant energy $E_{res}$ can only pass through the tunneling barrier when the chemical potentials of left and right electrodes align approximately with each other.

The transmission profiles, as displayed in Fig. 2 reflect the charge transport behavior of various molecular junctions under study. Since the resonant peaks at the lowest unoccupied molecular orbital (LUMO) are closer to the Fermi level, the electronic conductances appear to be dominated by it. Further, the transmission lineshape takes the shape of Breit–Wigner resonances in each case. In molecular junctions having Cr-Cr-Cr, Cr-Ru-Cr and Ru-Ru-Ru as the respective metal-string, we observe two closely peaked Breit–Wigner (BW) resonances, as shown in Fig. 2 (a,c,e) while only one such peak appears for Cr-Cr-Ru and Cr-Ru-Ru based junctions (see Fig. 2(b,d)). Taking into account all the frontier orbitals,  the transmission lineshape may be approximated as\cite{rincon2016}

\begin{equation}
T(E)=\sum_{n=1}^{2}\frac{4\Gamma_{n}^{L}\Gamma_{n}^{R}}{(E-E_n)^2+4{\bar{\Gamma}_{n}}^2},
\end{equation}  

where $\Gamma_{L}$ ($\Gamma_{R}$) represents the coupling of left (right) electrodes with the frontier orbitals of the respective single-molecule so that the average coupling is given by ${\bar{\Gamma}_{n}}=(\Gamma_{n}^{L}+\Gamma_{n}^{R})/2$; $E_n$ is the energy at which the resonant peak occurs  for the $n$-th BW resonance. This simple model, as depicted in Eq.(6), allows us to understand that the steepness in transmission near the Fermi level can yield better thermoelectric response. The calculated coupling strength for each of the metal-string junctions is mentioned in the individual subplots of Fig. 2 ($\Gamma_1$ and $\Gamma_2$ correspond to the first and second resonant peak respectively). The onset of highly intense transmission peaks in Cr-Ru-Cr junctions may be attributed to symmetric coupling formed at the electrode-molecule interface ($\Gamma_L/\Gamma_R=1$), while the resonance seems to be very weak in the case of Cr-Cr-Ru junctions due mainly to asymmetric coupling ($\Gamma_L\lll\Gamma_R$). Table 1 furnishes the coupling form at the left and right interfaces of the metal-string junctions. Furthermore, the respective plots of \textit{local density of states} (LDOS) indicate that for certain junction configurations (see Fig. 2 (b,d)), one of the two renormalized molecular levels forming the present DQD system possesses a  weakly localized state, acting thereby as a non-resonant state. 

\begin{table*}[htb!]
	\caption{ Coupling at the left/right electrode-molecule interface ($\Gamma_L/\Gamma_R$) as well as the average coupling ($\Gamma_1/\Gamma_2$) associated to each resonant peaks(1, 2) in the transmission profile of the metal-string junctions within the same bias window.}
	
	\label{tbl:example}
	\begin{tabular*}{\textwidth}{@{\extracolsep{\fill}}|c|ccc|ccc|}
		\cline{1-7}
		Metal-&\multicolumn{3}{c|}{\textbf{Peak 1}}& \multicolumn{3}{c|}{\textbf{Peak 2}}\\
		\cline{2-7}
		string& $\Gamma_L$&$\Gamma_R$&$\Gamma_1$&$\Gamma_L$&$\Gamma_R$&$\Gamma_2$\\
		\cline{1-7}
		Cr-Cr-Cr&	0.58&	0.02&	0.30&	3.39&	0.06&	1.70\\
		Cr-Cr-Ru & 0.01 & 3.29 & 1.60&&&\\
		Cr-Ru-Cr &	0.14 &	0.14 &	0.14 &	2.50 &	2.50 &	2.50\\
		Cr-Ru-Ru &	0.12&	2.58 &	1.40 &&&\\
		Ru-Ru-Ru &	3.12 &	0.10&	1.60&	4.00&	0.08&	2.00\\
		\cline{1-7}
	\end{tabular*}
\end{table*}  

In the ruthenium complexes, the contribution of $\pi$ component to the net bonding often leads to shorter Ru-Ru bond lengths and a LUMO with anti-bonding $\pi^*$ orbital (a second-order \textit{Jahn–Teller} effect)\cite{Mohan2012}. A strong coupling between $\pi^*$ and $\sigma_{nb}$ orbitals eventually results in the significant bending as high as about $170^o$ in the tri-ruthenium (Ru-Ru-Ru) metal-string\cite{Mohan2012,Niskanen2012}. This leads to the cancellation of spin polarization which is responsible for the introduction of band gap around the Fermi level. On the other hand, bending in the metal core of Ru-Ru-Ru chain incorporates the asymmetry in the Ru-N bond lengths ranging from 1.9 to 2.1 $\AA$, which are in line with the observations made by Mohan \textit{et al}\cite{Mohan2012}. Additionally, strong delocalisation in these metal-string complexes generates broad $\pi$ bands which are not easily perturbed by the application of an external  electric field. Such peculiar behavior of Ru makes it distinct from its first-row counterparts such as Cr, Co etc. Dominant transport channel for tri-ruthenium complexes are mainly from $\pi^*$ orbitals while tri-chromium complexes are dominated by weakly conducting $\sigma_{nb}$ orbitals\cite{Mohan2012}. The mixing of Cr and Ru in the metal-string gives rise to distinct transmission profile due to contributions from $\pi^*$ as well as $\sigma_{nb}$ orbitals towards the charge transport, as demonstrated in Figs. 2(b-d). We find that both structure and conductance are intertwined with each other, as indicated by Walsh’s rule\cite{Walsh1953}, since the structural distortion comes mainly from the perturbation of frontier orbitals. The resonant peaks thus tend to move away from the Fermi level upon systematic rutheniation. 

Orbital characters at the contacts made by respective molecular moieties with electrodes at the left and right interfaces are portrayed in Fig.3. It shows that $\sigma$ and $\pi$ channels can be formed at the contacts  due mainly to the interaction between $s$ or $d$ orbitals from the Au adatom and $p$ orbitals from the S atom. These conducting channels facilitate the charge transport across various metal-string complexes. Understanding the essence of phonon dynamics in order to manipulate the thermal properties is also important since thermoelectricity is a material property that signifies the conversion efficiency of thermal into electrical energies. Usually in 1D nanostructures, the phonon dispersion relations are modified significantly due to confinement in the other two directions. It further modifies the group velocity as well as the phonon lifetime due to changes in the phonon-phonon interaction and strong boundary scattering. The dominating vibrational modes present in the molecular moieties turn out to be out-of-plane $stretching$ ($\nu$), out-of-plane $bending$ ($\delta$), out-of-plane $16b$ and in-plane $6b$ modes, as demonstrated in Figs. 3(a-e). The two degenerate pairs of $16b$ and $6b$ modes, having $e_{2g}$ and $e_{2u}$ symmetry, depict a close resemblance with the vibrations of $pyridyl$ rings present in metal-string complexes, according to the Wilson-Varsanya terminology (WVT)\cite{Varsanyi1974,Varsanyi1969}.

Fig. 4 illustrates the temperature effect to the thermoelectric response such as conductance ($ \sigma $), thermopower ($S$), and the electronic components of thermal conductance ($\kappa_{el}$) for various metal-string molecular junctions under study. Symmetric metal-string junctions such as  Cr-Ru-Cr and Ru-Ru-Ru show high conductances as compared to the asymmetric ones (see Fig. 4(a)). However, in the case of $tri-ruthenium$ based molecular junctions, the electronic and thermal conductances tend to increase gradually with temperature. The discrete molecular energies straddle the Fermi level of the gold electrodes, contributing thereby to the conductance variation. When metal-string complexes form the molecular junctions, magnitude of thermopower as high as 172 $\mu V/K$ can be achieved at room temperature, which is higher than the one reported by Taniguchi and his co-workers\cite{Tsutsui2009}, who attained a thermopower value of 120 $\mu$V/K through mechanical stretching of benzene-di-thiol (BDT) junctions. Besides, Cr-Cr-Cr metal-string junctions are observed to show good themopower as well within the temperature range from 170 to 350 K eventually with high power factor. As observed by Kl$\ddot{o}$ckner \textit{et al}\cite{klockner2017}, pure electron contribution towards the thermal conductance in single molecular junction cannot be ignored. Thus \textit{tri-ruthenium} based junctions, which are purely symmetrical, yield high electronic thermal conductance as depicted in Fig. 4(c). On the other hand, devices made out of single molecular junctions often exhibit strong Coulomb interactions leading to the violation of the Wiedemann-Franz (WF) law\cite{Wierzbicki2011,elke2017,burkle2015} defined as $L_0=(\kappa_{0}h)/(e^2 T)=(\pi^2/3)(K_B/e)^2=2.44 \times 10^{-8}$ W$\Omega K^{-2}$ where $\kappa_{0}$ = $(\pi^2/3)(K_B^2T/h)$ is the quantum of thermal conductance. The non-linear behavior of ($\kappa_{el}/\sigma$) vs. $T$ as seen in Fig. 4(d) provides a clear indication of significant variations in the Lorenz number due mainly to interference from phonons. 

Fig. 5(a) shows the phonon transmission profiles for all the systems under study. 
The phonon thermal conductance for the metal-string junctions saturate around Debye temperature of gold (i.e. 170 K) as observed in Fig. 5(b). At low temperature (T), the phonon thermal conductance become less sensitive to T because of the partial occupancies of the high-lying modes. But above the Debye temperature, all the vibrational modes are thermally occupied within the transport regime. The thermoelectric \textit{figure of merit ($ZT$)} as a function of energy and temperature are displayed respectively in Fig. 5(c) and Fig. (d) where $chromium$ based metal-string junctions are observed to show good thermoelectric properties, even though a higher value is shown by non-rutheniated Cr-Cr-Cr junctions. These junctions exhibit an increase in $ZT$ within 170 K, as suggested by Fig. 5(d). The increase in $ZT$ for the metal-string molecular junctions may be attributed to the  high reduction in the total thermal conductance, due to the quantum size effect\cite{Dresselhaus1993} as well as large mismatch of vibrational spectra between the nanowire-based electrodes and discrete molecules. However, the thermopower gets enhanced mainly by the point contacts made by the respective molecule and electrode through the gold adatoms\cite{Wierzbicki2011,ROY2018}. Our calculated values of thermopower and \textit{figure of merit ($ZT$)} at 300 K for all the metal-string based molecular junctions under study are shown in Fig. 6 (a-b). The negative (positive) sign for the  thermopower indicate dominant charge transport where electrons (holes) tunnel through the LUMO (HOMO). Out of all the rutheniated and non-rutheniated $chromium$-based molecular junctions, it turns out that Cr-Cr-Cr demonstrates better room-temperature thermoelectric property with $ZT$ being close to  2. However, rutheniated junctions display much less thermoelectric response due to high phonon thermal conductance (see Fig. 5(c)).   

\section{Conclusions}
For molecular junctions comprising metal-string complexes, we have found that the LUMO-mediated transmission is primarily responsible for the charge transport behavior in presence of strong electrode-molecule coupling. Our \textit{first-principles} quantum transport analysis  further affirms that metal-string junctions can exhibit strong coulomb interactions leading to violation of the Wiedemann-Franz (WF) law. Resonant enhancement in the thermoelectric efficiency is attained through alteration of inter-dot electrostatic interactions which can further be controlled by systematic rutheniation in tri-nuclear metal-string complexes. Out of all the metal-string complexes, Cr-Cr-Cr based molecular junctions can only have the thermopower as high as 172 $\mu V/K$, with the thermoelectric \textit{figure of merit} ($ZT$) of about 2. Metal-string based single  molecular devices can thus potentially serve as good thermoelectric devices with improved efficiency and stability.

\section{Acknowledgement}

We acknowledge the DST Nano Mission, Govt. of India for the necessary financial support, $via$ Project No. SR/NM/NS1062/2012. TRR acknowledges the Council of Scientific and Industrial Research(CSIR), New Delhi,  for providing the senior research fellowship (File no. 09/1045(0017)2K18). We are also thankful to SRM-HPCC, SRM Institute of Science and Technology for facilitating the high-performance computing.

\section*{}

\bibliography{rebeda_arxiv_ic2}

\newpage
\begin{figure*}
	\centering
	\includegraphics[scale=1.2]{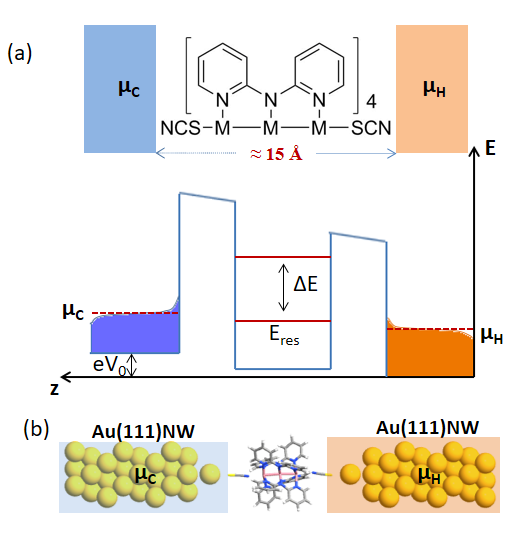}
	\caption{(a) Schematic of metal-string molecular junctions comprising $[M-M-M](dpa)_4(NCS)_2$ (where $M \equiv Cr; Ru$), coupled to cold and hot electrode reservoirs having respective electrochemical potentials $\mu_C$ and $\mu_H$. A double quantum dot (DQD) model is furnished subsequently, where $\Delta E$ signifies the energy difference between the two quantum dots. When $\Delta E \gg K_BT$, a single resonance with energy $E_{res}$ dominates the transmission. (b) A two-probe systems consisting of a metal-string sandwiched between two gold electrodes of Au(111) nanowire type.}
	\label{fgr:figure1col}
\end{figure*}

\begin{figure*}
	\centering
	\includegraphics[scale=0.6]{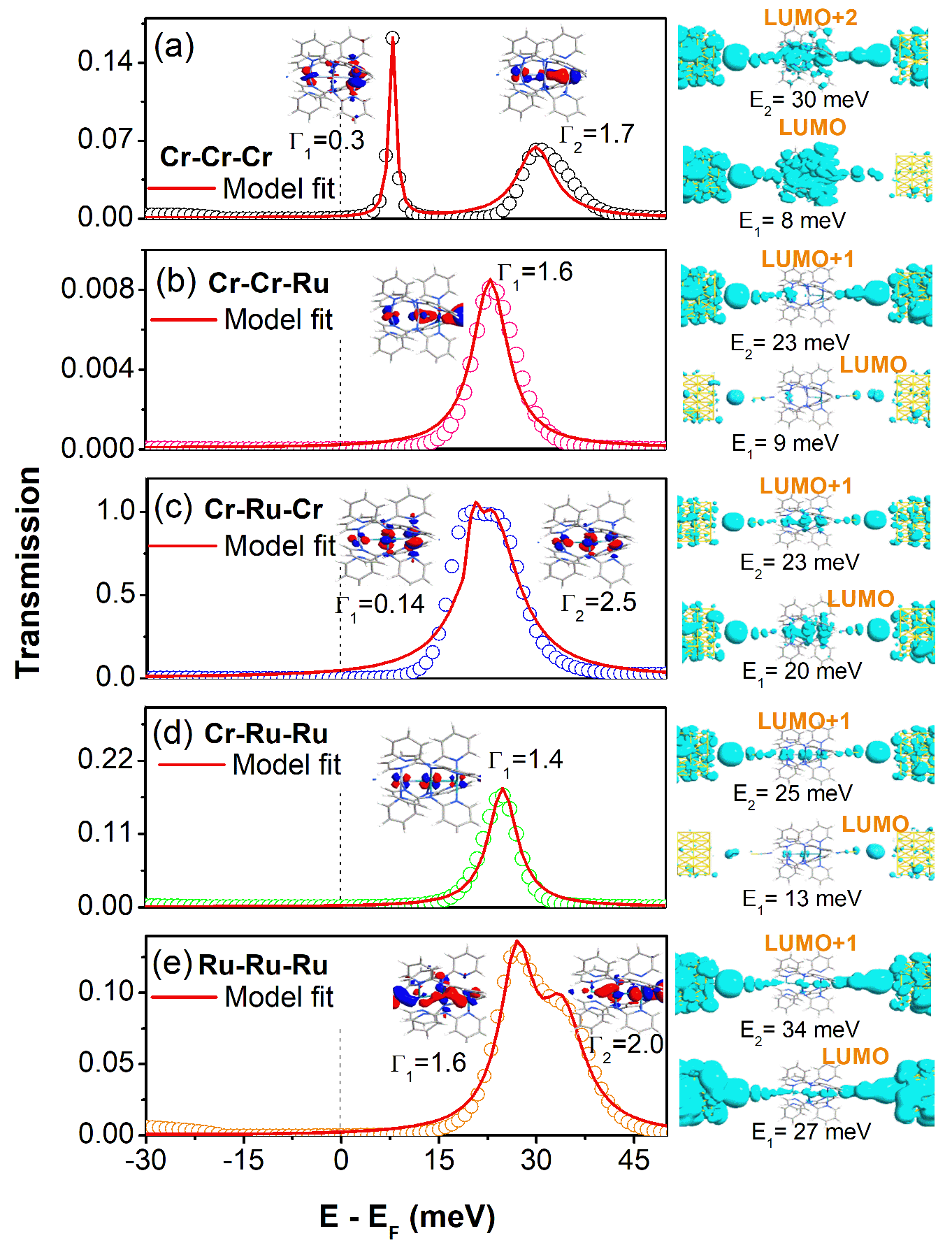}
	\caption{Transmission profiles for each of the molecular junctions having (a) Cr-Cr-Cr, (b) Cr-Cr-Ru, (c) Cr-Ru-Cr, (d) Cr-Ru-Ru and (e) Ru-Ru-Ru as the respective metal-string based molecular moiety. The lineshapse were fitted (red line) using the DQD model (see the text) as given in Eq. (6). The respective values of the level broadening ($\Gamma_n$, for $n \equiv 1; 2$) associated with each Breit-Wigner peak are mentioned along with their respective eigenstates plots. The respective plots for device local density of states (LDOS) corresponding to renormalized molecular levels of the frontier orbitals are also depicted in the right panel.}
	\label{fgr:figure1col}
\end{figure*}

\begin{figure*}
	\centering
	\includegraphics[scale=0.81]{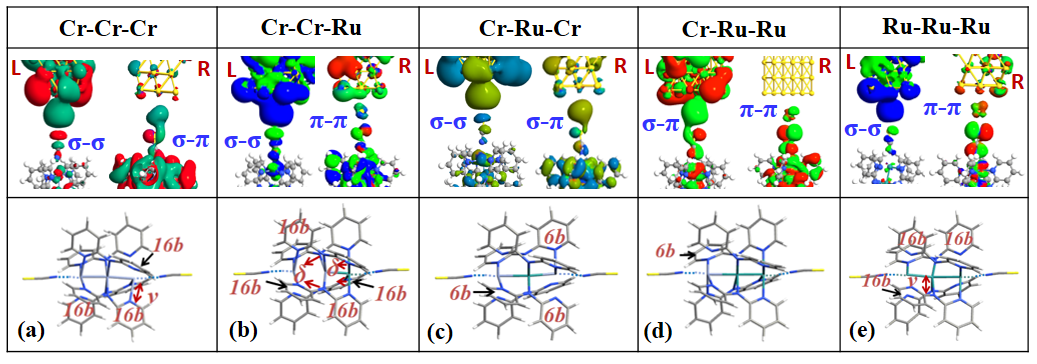}
	\caption{Transmission eigenstates for (a) Cr-Cr-Cr, (b) Cr-Cr-Ru, (c) Cr-Ru-Cr, (d) Cr-Ru-Ru and (e) Ru-Ru-Ru based molecular junction indicating the type of orbital contacts which may be formed at the electrode-molecule interfaces. In the bottom panel are shown the nature of dominant vibrational modes persisting in respective molecular moieties as a result of the phonon emission. }
	\label{fgr:figure1col}
\end{figure*}

\begin{figure*}
	\centering
	\includegraphics[scale=0.9]{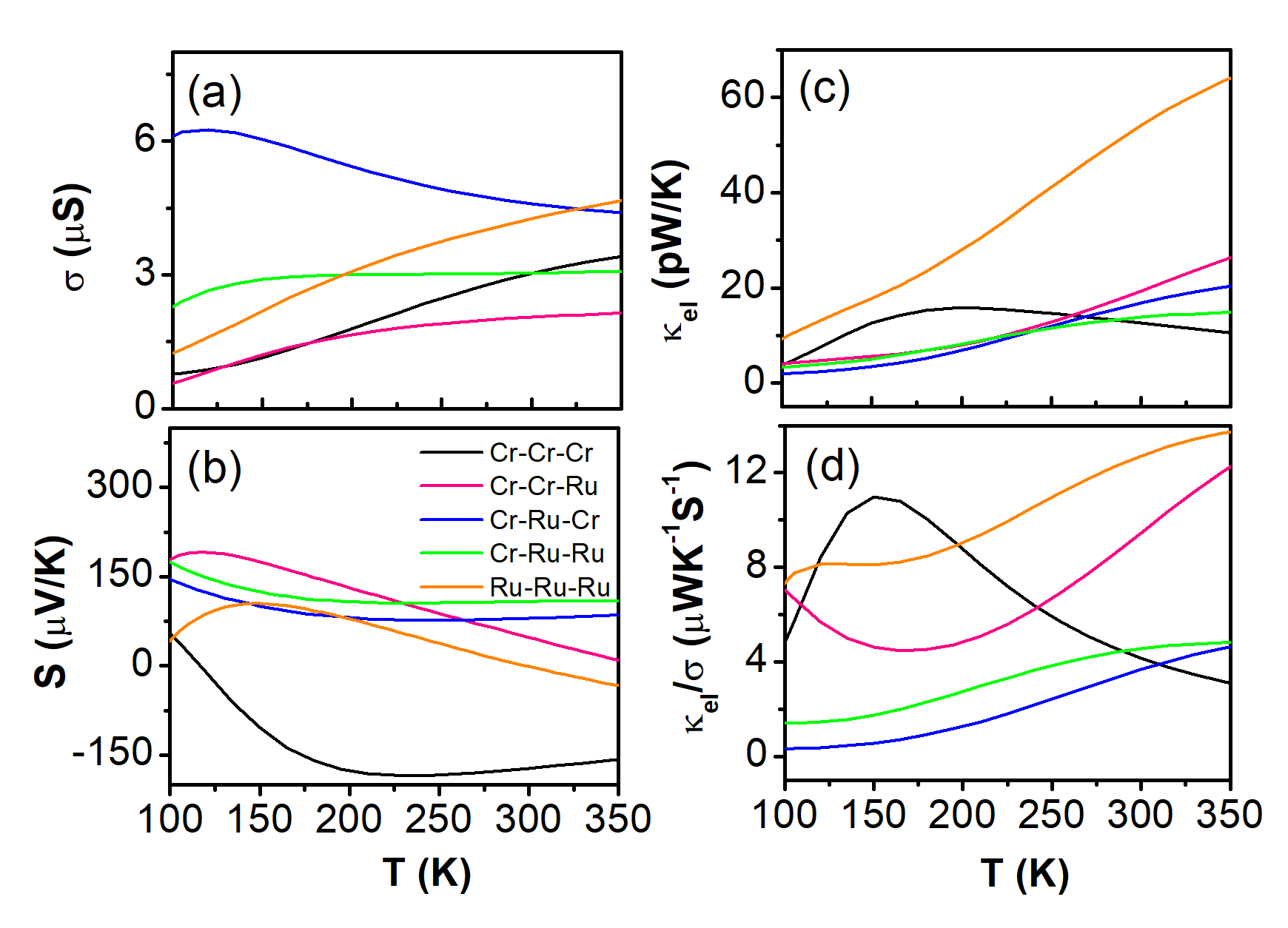}
	\caption{Plots showing various thermoelectric responses such as (a) electronic conductance, $\sigma$, (b) thermopower, S, (c) electronic component of the thermal conductance, $\kappa_{el}$, and (d) electronic component of the thermal conductance to electronic conductance ratio, all as a function of temperature. A nonlinear behavior is observed here, as  $\kappa_{el}/\sigma T \neq L_0$, with $L_0$ being the Lorenz number.}
	\label{fgr:figure1col}
\end{figure*}

\begin{figure*}
	\centering
	\includegraphics[scale=0.61]{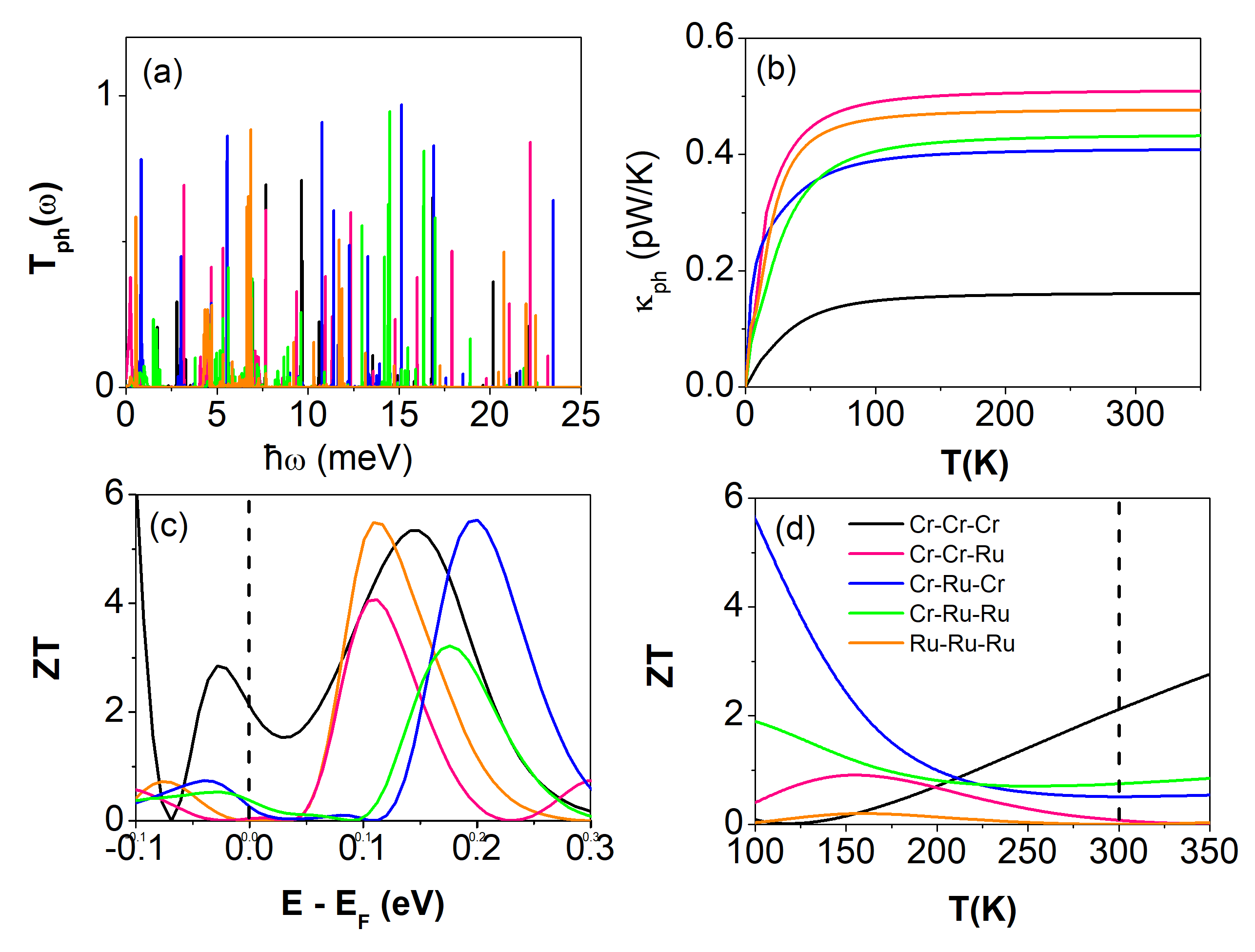}
	\caption{(a) Phonon transmission function, (b) temperature dependence of the phonon contribution to thermal conductance ($\kappa_{ph}$), (c) thermoelectric \textit{figure of merit} ($ZT$), as a function of energy (relative to the Fermi level) at 300 K,  and (d) $ZT$ as a function of temperature for various metal-string molecular junctions under study.}
	\label{fgr:figure1col}
\end{figure*}

\begin{figure*}
	\centering
	\includegraphics[scale=0.5]{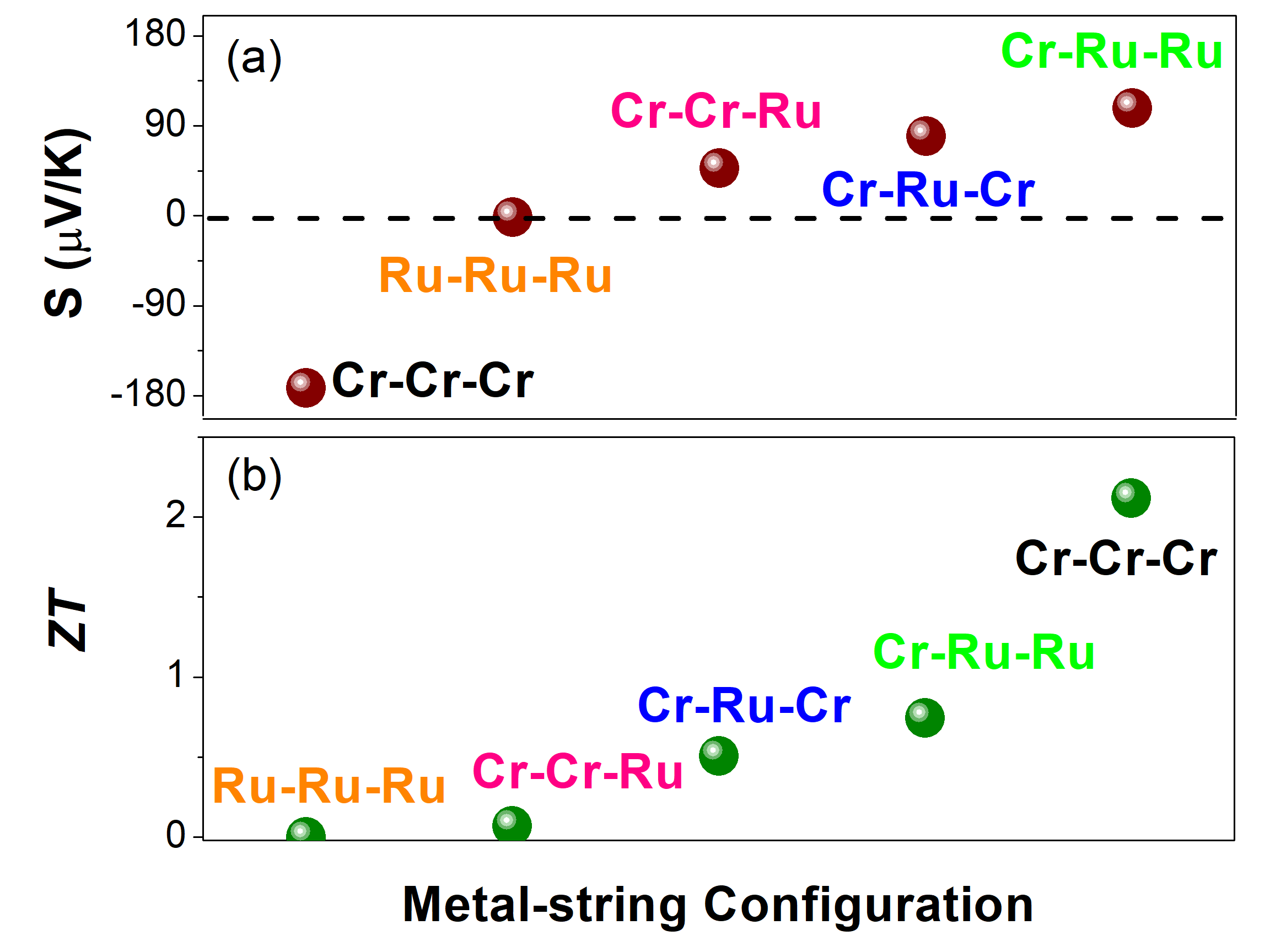}
	\caption{Field plots showing the room temperature (a) thermopower and (b)  thermoelectric \textit{figure of merit}, $ZT$, in association with various metal-string molecular junctions under study.}
	\label{fgr:figure1col}
\end{figure*}

\end{document}